\documentclass{PoS}

\usepackage{color} 
\usepackage{graphicx} 
\usepackage{multirow}
\usepackage{hhline}
\usepackage{tikz}

\usepackage{hhline}
\usepackage{tikz}
\usepackage{dutchcal}
\usepackage{subfigure}
\usepackage{transparent}

\usepackage{lmodern}

\usepackage[T1]{fontenc}
\usepackage[latin9]{inputenc}
\usepackage{booktabs}
\usepackage{mathrsfs}
\usepackage{amsmath}
\usepackage{amssymb}
\usepackage{graphicx}

\title{BSM Physics from Enlarged Gauge Symmetry: the 331 Model, a case of study}

\ShortTitle{The 331 Model, a case of study}

\author{\speaker{Antonio Costantini}\\
        INFN - Sezione di Bologna, Via Irnerio 46, 40126 Bologna, Italy\\
        E-mail: \email{antonio.costantini@bo.infn.it}}

\abstract{We discuss the most relevant features of a BSM model with extra gauge symmetry, the so called 331 model. The gauge group is $SU(3)_c\times SU(3)_L \times U(1)_X$ impling the presence of extra gauge bosons, both charged and neutral, as well as extra/exotic fermions and an enlarged scalar sector. We present the relevant phenomenology of doubly-charged gauge bosons, which are a distinctive feature of a version of the 331 model, and discuss the role of BSM phenomenology as a tool for testing GUT (inspired) theories.}

\FullConference{
European Physical Society Conference on High Energy Physics - EPS-HEP2019 -\\
			10-17 July, 2019\\
			Ghent, Belgium}

\begin{document}

\section{Introduction}
In spite of its success in the explanation of the Elementary Particle's dynamics, the Standard Model (SM) seems to be unable to sustain certain experimental evidences that do not fit in its current formulation. Moreover, there are theoretical problems that suggest that the SM should be thought as a low-energy theory, embedded in a larger model at scales higher than the electroweak one. We will limit ourselves in list some of the theoretical and phenomenological issues of the SM, without any pretension of completeness. 

One of the most celebrated achievements of the SM, the discovery of the Higgs boson made by the ATALS and CMS collaboration of the LHC in 2012 \cite{Aad:2012tfa,Chatrchyan:2012xdj}, poses the SM itself in some problems. There is no symmetry in the SM that protects the Higgs boson mass against quantum corrections, hence large fine tuning in required to have $m_h = 125.09$ GeV, as measured at the LHC. 

In the SM neutrinos are massless neutral leptons. This is clearly in contrast with experimental evidences of neutrino oscillations \cite{Fukuda:1998fd,Araki:2004mb}. Although there are no conclusive measurements on the mass of each neutrino flavour and it is still not clear if neutrinos are Dirac or Majorana fermions, the measured flavour oscillations imply that they 
are not massless.

Experimental evidence of dark matter \cite{Rubin:1978kmz,Rubin:1980zd,Persic:1995ru}, know since more than four decades, are unable to be understood within the context of the SM. A stable massive particle, which is one of the possibility of the dark matter, does not exist in the original formulation of the SM.

As a final remark, let us remind that there is no explanation in the SM for $n_{f_L}=n_{f_Q}=3$. The electroweak precision measurements on the $Z$ resonance \cite{ALEPH:2005ab} are in agreement with $n_{f_L}=3$ but the SM does not provide any arguments for $n_{f_L}=n_{f_Q}$. We will come back to this issue in the next section. 

\section{The 331 Model}
Let us now consider an extension of the SM obtained starting from a larger gauge group. The gauge symmetry of the model is
\begin{equation}
SU(3)_c\times SU(3)_L\times U(1)_X
\end{equation}
Clearly the $U(1)$ factor is not the SM hypercharge by itself. In fact $SU(3)_L$ has two diagonal generators hence the electromagnetic charge operator is
\begin{equation}
\mathbb Q \equiv \mathbb Y + \mathbb T_3 = \beta^{em} \mathbb T_8 + \mathbb X + \mathbb T_3\nonumber
\end{equation}
The 331 model is actually a class of models, parametrized by the possible values of $\beta^{em}$. Although at this stage this is a free parameter, we will see that in order to fulfil some basic requirements, such as the quantization of the electromagnetic charge, $\beta^{em}$ can take only few values.
 
The enlarged gauge symmetry implies that we must extend the matter content of the model. Scalars and fermions are in fact arranged in triplets, being in the fundamental (or anti-fundamental) representation of $SU(3)$. The 331 model treats differently quarks and leptons. In the quark sector we have
\begin{equation}\label{eq:quarks}
Q_1=\left(
\begin{array}{c}
u\\
d\\
D
\end{array}
\right),\quad Q_2=\left(
\begin{array}{c}
c\\
s\\
S
\end{array}
\right),\quad Q_{1,2}\in(\mathbf 3, \mathbf 3, X_{Q_{1,2}})
\end{equation}
\begin{equation}
Q_3=\left(
\begin{array}{c}
b\\
t\\
T
\end{array}
\right),\quad Q_3\in(\mathbf 3,\bar{\mathbf 3}, X_{Q_3})
\end{equation}
whereas the leptons are arranged as
\begin{equation}\label{eq:leps}
L=\left(
\begin{array}{c}
l\\
\nu_l\\
E_l
\end{array}
\right),\quad L\in(\mathbf1, \bar{\mathbf 3}, X_{L}),\quad l=e,\ \mu,\ \tau
\end{equation}
We have not written explicitly the $U(1)_X$ quantum number for the fermions.
We notice that the first two families of quarks belong to the fundamental of $SU(3)_L$ whereas the third one belong to the anti-fundamental. In the lepton sector there is not such a difference, all the flavour belong to the same representation of $SU(3)_L$. We explain the reason for this different treatment in a moment. Before that, we remark that we have not written the right-handed fermions, which are of course present in the 331 model.

\begin{figure}[t!]
\centering
\includegraphics[scale=.75]{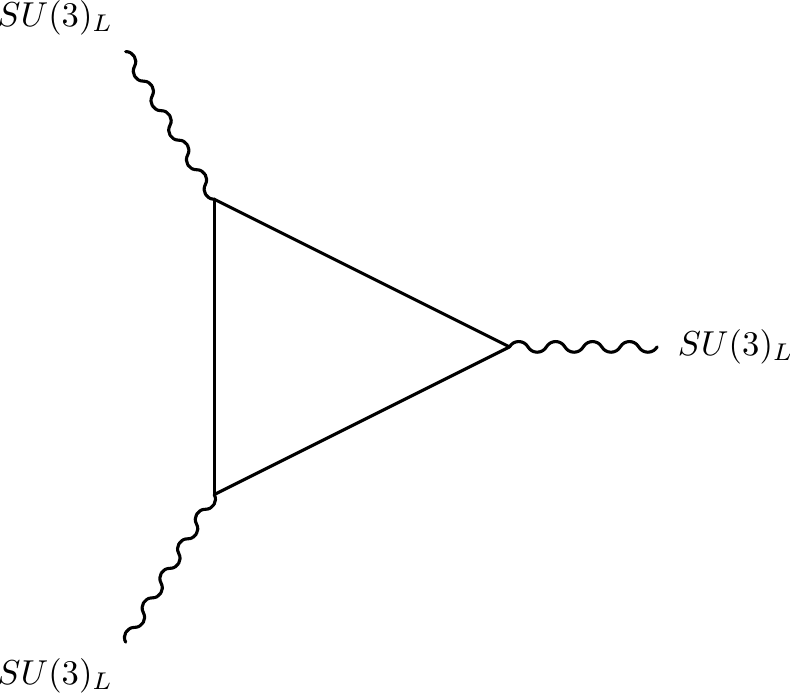}
\caption{Schematic representation of the loop diagram involved in the $(SU(3)_L)^3$ anomaly.}\label{fig:anomaly}
\end{figure}

The reason for the asymmetry in $SU(3)_L$ representation of quarks and leptons lies in the anomaly cancellation within the 331 model. In Figure~\ref{fig:anomaly} is shown schematically the loop diagram involved in the $(SU(3)_L)^3$ anomaly. The fermion lines correspond to quarks and leptons charged under $(SU(3)_L)^3$ hence precisely the ones in Eqs.~(\ref{eq:quarks})-(\ref{eq:leps}). This anomaly vanish because of the equal number of fundamental and anti-fundamental representations, once the colour multiplicity of the quarks has been taken into account. Anomaly cancellation in the 331 model then forces
\begin{equation}
n_{f_L}=n_{f_Q} = 3\kappa
\end{equation}
giving an explanation for the equal number of quark and lepton families. Anomaly cancellation in the 331 model happens among the three families of fermions, conversely from the SM case.

\begin{table}[t!]
\centering
\renewcommand\arraystretch{1.2}
\begin{tabular}{c | c |c |c| c| c }\hline
 \textrm{particle}& $Q(\beta^{em})$ & $\beta^{em}=-\frac{1}{\sqrt{3}}$ & $\beta^{em}=\frac{1}{\sqrt{3}}$ & $\beta^{em}=-\sqrt{3}$ & $\beta^{em}=\sqrt{3}$ \\ \hline \hline
$D,S$ & $\frac{1}{6}-\frac{\sqrt{3}\beta^{em}}{2}$ & $\frac{2}{3}$ & $-\frac{1}{3}$ & $\frac{5}{3}$ & $-\frac{4}{3}$ \\ \hline
$T$ & $\frac{1}{6}+\frac{\sqrt{3}\beta^{em}}{2}$ & $-\frac{1}{3}$ & $\frac{2}{3}$ & $-\frac{4}{3}$ & $\frac{5}{3}$ \\ \hline
$E$ & $-\frac{1}{2}+\frac{\sqrt{3}\beta^{em}}{2}$ & $-1$ & $0$ & $-2$ & $1$ \\ \hline
$V$ & $-\frac{1}{2}+\frac{\sqrt{3}\beta^{em}}{2}$ & $-1$ & $0$ & $-2$ & $1$ \\ \hline
$Y$ & $\frac{1}{2}+\frac{\sqrt{3}\beta^{em}}{2}$ & $0$ & $1$ & $-1$ & $2$ \\ \hline
\end{tabular}
\caption{\it Electric charges of new particles for different choices of $\beta^{em}$.}
\label{tbl:charge}
\end{table}

The scalar sector consist in three triplets
\begin{equation}
\chi=\left(
\begin{array}{c}
\chi^{A}\\
\chi^B\\
\chi^0
\end{array}
\right)\in(1,3,X_\chi),\quad\rho=\left(
\begin{array}{c}
\rho^+\\
\rho^0\\
\rho^{-B}
\end{array}
\right)\in(1,3,X_\rho),\quad\eta=\left(
\begin{array}{c}
\eta^0\\
\eta^-\\
\eta^{-A}
\end{array}
\right)\in(1,3,X_\eta)
\end{equation}
Here again we have left unspecified their $U(1)_X$ quantum numbers. The electromagnetic charges of the A- and B-states are $Q^A=\frac{1}{2}+\frac{\sqrt3}{2}\beta^{em}\;,\,Q^B=-\frac{1}{2}+\frac{\sqrt3}{2}\beta^{em}$. They depend of course on $\beta^{em}$ as the electric charge of the extra fermionic degrees of freedom. The neutral component of each triplet can take a vacuum expectation value (vev). In this way the gauge symmetry is spontaneously broken, in complete analogy with the electroweak symmetry breaking in the SM. The spontaneous symmetry breaking chain is
\begin{equation}
SU(3)_L\times U(1)_X  \xrightarrow{v_\chi} \;SU(2)_L\times U(1)_Y  \xrightarrow{v_\rho,\,v_\eta}\; U(1)_{em}
\end{equation}

After the first breaking, when $\chi$ gets vev, 3 gauge bosons became massive. They are $Z^\prime_\mu,Y_\mu^{\pm A}, V_\mu^{\pm B}$. The $Z^\prime_\mu$ is a mixture of $X_\mu$ and $W_\mu^8$ whereas
\begin{equation}
Y_\mu^{\pm A}=\frac{1}{\sqrt2}(W^4_\mu \mp i\, W^5_\mu)\,,\;
V_\mu^{\pm B}=\frac{1}{\sqrt2}(W^6_\mu \mp i\, W^7_\mu)
\end{equation}
The mass of $Z'_\mu$ is given by \cite{Buras:2012dp}
\begin{equation}\label{eq:mzp}
M^2_{Z^\prime}=\frac{g^2 v_\chi^2 \,\cos\theta_W}{3(1-(1+(\beta^{em})^2\sin^2\theta_W))}(1+\cdots)
\end{equation}
From Eq.~(\ref{eq:mzp}) we obtain that, given the value of the Weinberg angle, $|\beta^{em}|\leq\sqrt3$ if the $Z^\prime_\mu$ mass has to be positive definite. Moreover this parameter enters in the electromagnetic charge of the particles, which can be at most fractional. Hence we have $\beta^{em}=0,\pm1/\sqrt3,\pm2/\sqrt3,\pm3/\sqrt3$. In Table~\ref{tbl:charge} we give the electromagnetic charge for the extra fermions/gauge bosons in some interesting cases. 

Recently a phenomenological analysis has been done for the case $\beta^{em}=\sqrt3$. The phenomenological analysis concerns the production of \textit{dileptons} at the LHC \cite{Coriano:2018coq}. This version of the model in fact has the almost unique feature to accommodate a vector boson of charged $2$. Moreover, by inspection of Table~\ref{tbl:charge}, one recognize that the extra leptonic degree of freedom can be thought as the right-handed component of each charged lepton. Among the various version of the 331 model, that's the one in which the amount of extra fermions can be minimized.

\section{Same-sign Leptons Phenomenology}
Here we present the result of a phenomenological analysis \cite{Coriano:2018coq} that consider the production of a same-sign lepton pair. In the context of the SM this signature is absent and at the LHC the relevant background is given by $pp\to Z Z \to 2\ell^+2\ell^-$. The 331 model has an interesting scenario where the same-sign production of lepton pairs is allowed. In particular it can happens through a doubly-charged vector $Y^{\pm\pm}$ or a doubly-charged scalar $H^{\pm\pm}$. A scalar particle with charge two is predicted by various models of beyond-Standard-Model (BSM) physics, like models with triplets of $SU(2)_L$ or models with restored left-right symmetry at high energies. Conversely a doubly-charged vector is almost unique in BSM physics.

\begin{figure}[t!]
\centering
\mbox{\subfigure[]{\includegraphics[width=0.250\textwidth]{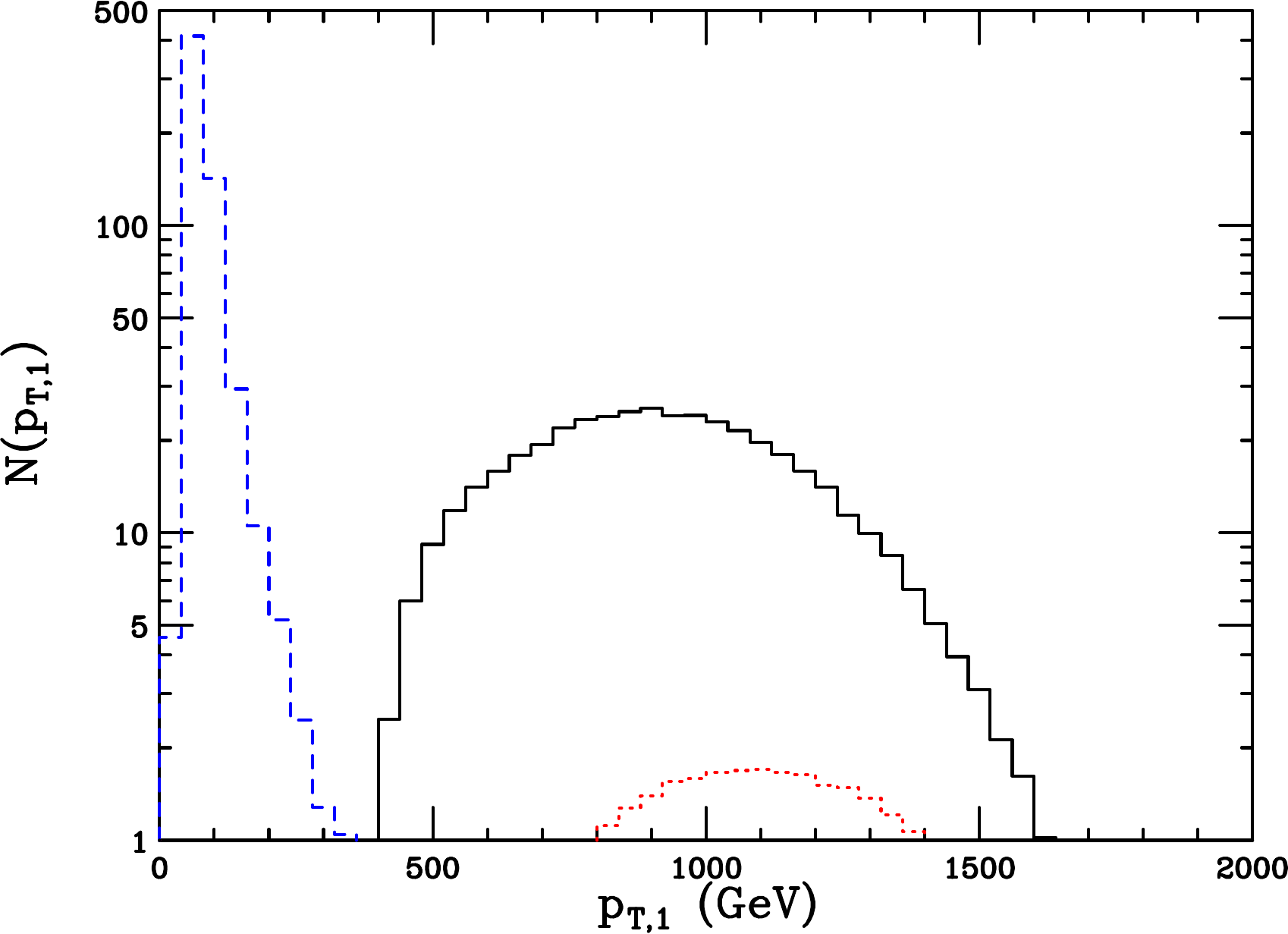}}
\subfigure[]{\includegraphics[width=0.250\textwidth]{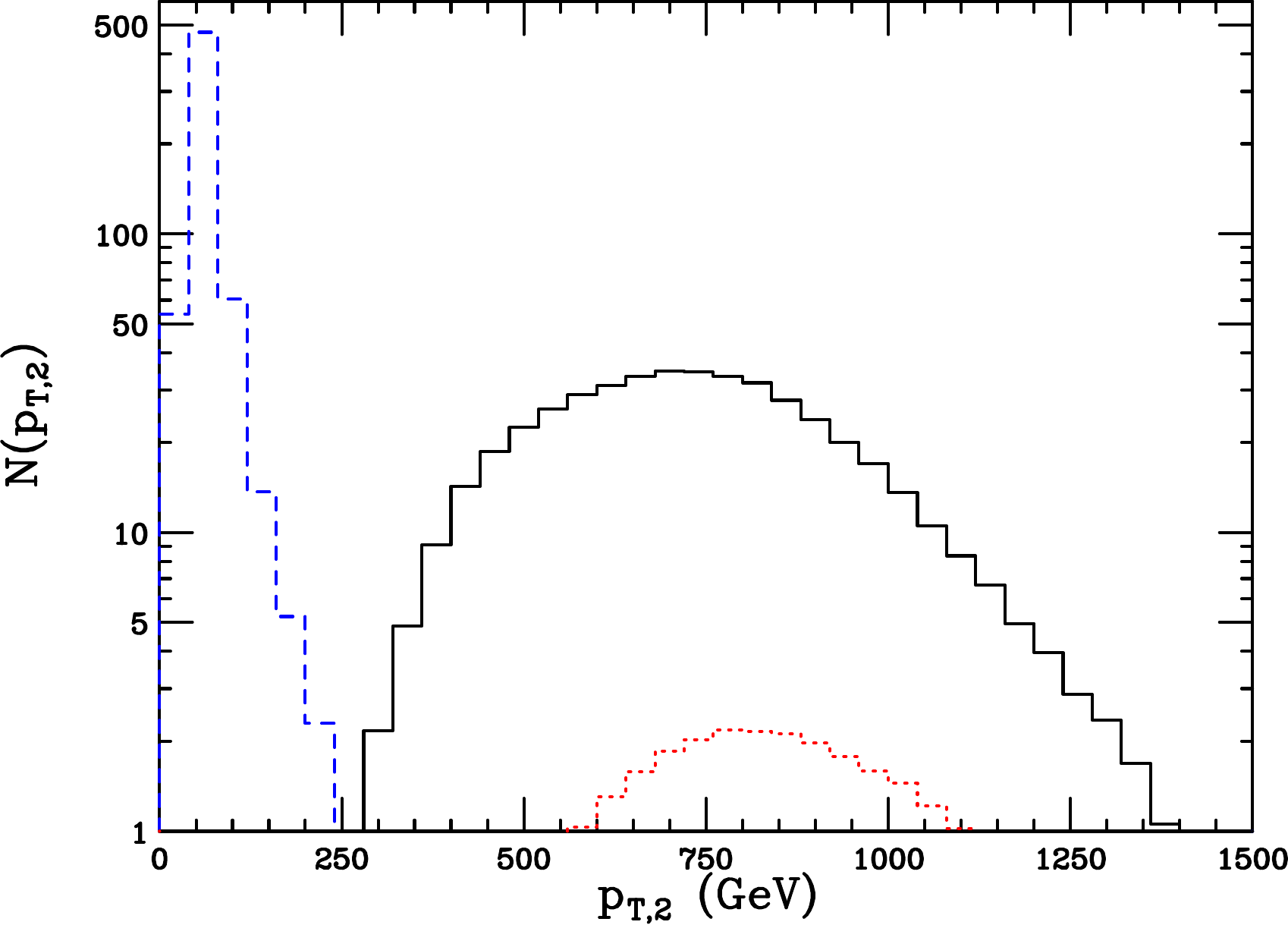}}
\subfigure[]{\includegraphics[width=0.250\textwidth]{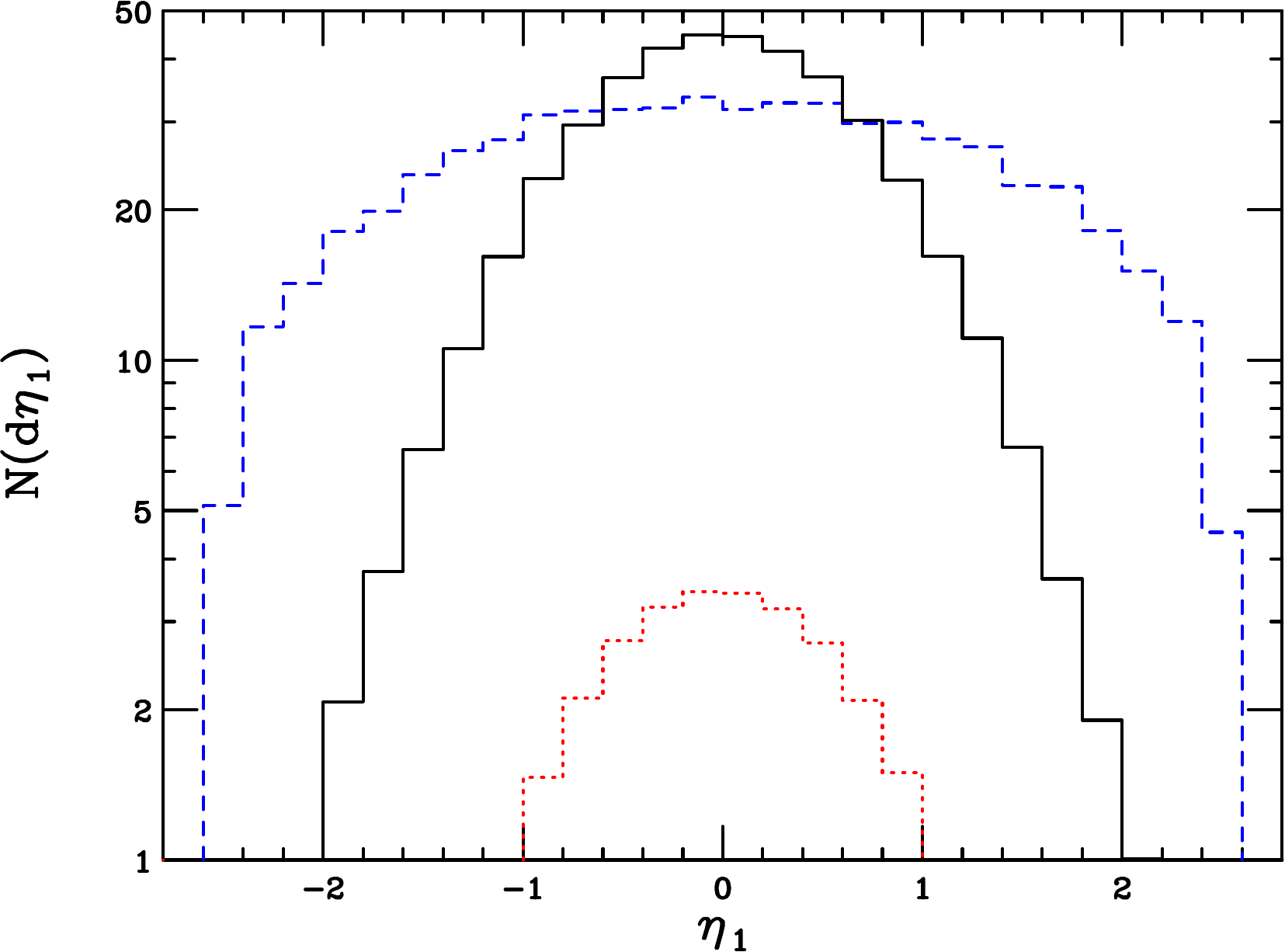}}
\subfigure[]{\includegraphics[width=0.250\textwidth]{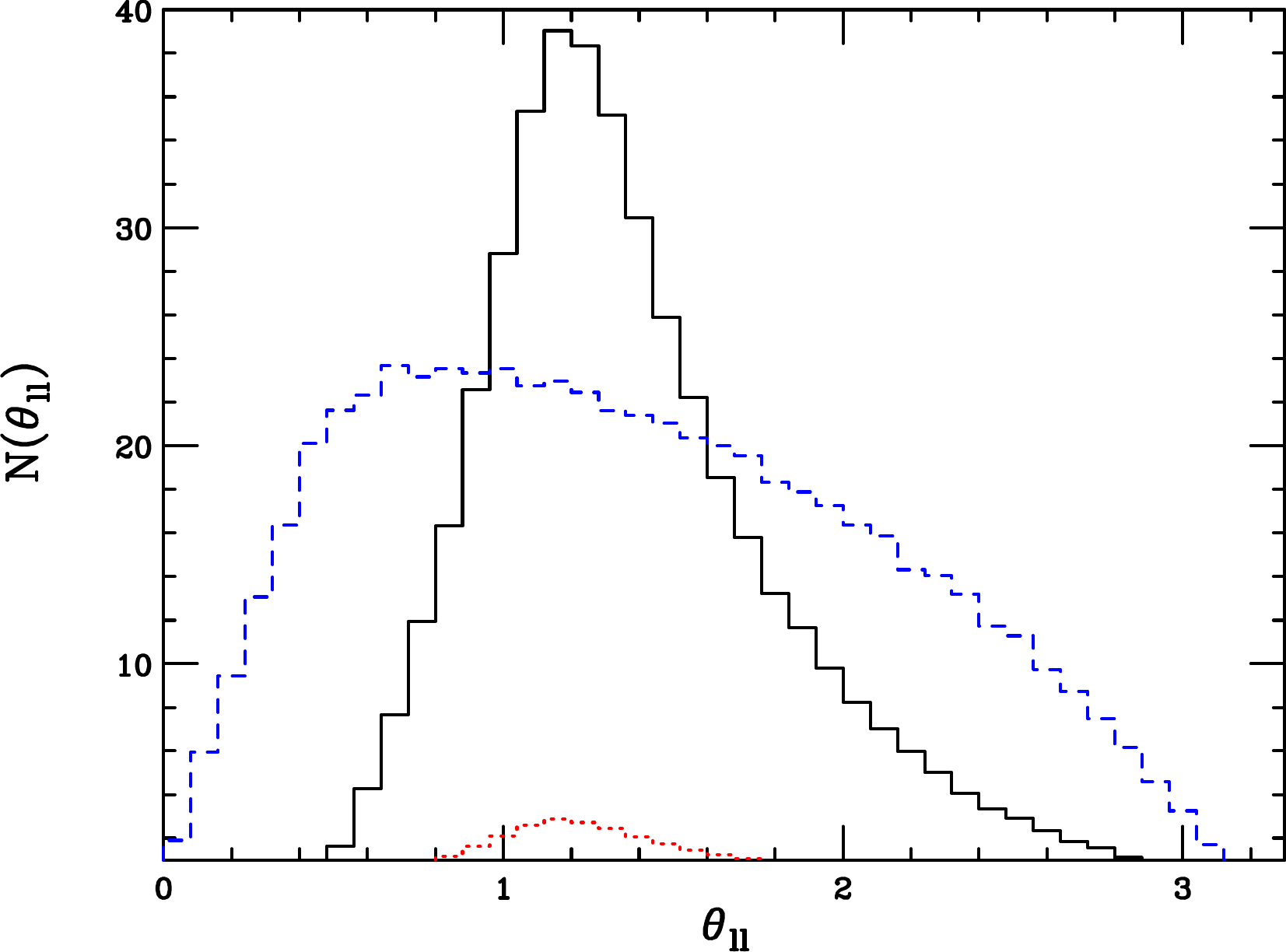}}}
\caption{Distribution of the $p_T$ of hardest (a) and next-to-hardest (b) lepton, $\eta$ of the leading lepton (c) and $\theta$ of the same-sign pair (d). Black solid is the $Y^{\pm\pm}$, red dotted is the $H^{\pm\pm}$ and blue dashed is the $Z$ pair respectively.}
\label{fig:distr}
\end{figure}

The phenomenological analysis presented in \cite{Coriano:2018coq} consider the process
\begin{equation}
pp\to B^{++}B^{--}\to 2\ell^+ 2\ell^- 
\end{equation}
at the LHC. Here $B^{\pm\pm}$ stands for either the scalar or the vector boson with charge two. The mass of these particles has been taken to be $m_{Y^{\pm\pm}}\simeq m_{H^{\pm\pm}}\sim870\; GeV$ and their branching fraction in leptons are $Br(Y^{\pm\pm}\to \ell^\pm \ell^\pm) = Br(H^{\pm\pm}\to \ell^\pm \ell^\pm)=1/3$.
The cross-section for the signal is $\sigma(pp\to YY\to 4\ell)\simeq
4.3$ fb and $\sigma(pp\to HH\to 4\ell)\simeq 0.3$ fb whereas the cross-section of the dominant background is $\sigma(pp\to ZZ\to 4\ell)\simeq 6.1$ fb, both at $\sqrt s=13$ TeV. In Figure~\ref{fig:distr} we give the distribution of the transverse momentum of the hardest (a) and next-to-hardest lepton (b), the rapidity of the leading lepton (c) and polar angle between the same-sign pair (d). By inspection of the distributions and/or calculating the significances for e.g. $300$ fb$^{-1}$ of integrated luminosity we conclude that the LHC will be sensitive to the spin-$1$ doubly-charged state predicted by the 331 model.

\section{Conclusions}
There are theoretical and phenomenological reasons to consider extensions of the SM. The search for physics BSM is usually driven by models that extend the field content or the gauge symmetry. The 331 model is a class of models that is able to predict the number of fermionic families by the anomaly cancellation. There are various versions, each of which has a different way to construct the SM hypercharge out of the generators of $SU(3)_L\times U(1)_X$. This makes the phenomenology of the 331 model very rich. This extension of the gauge symmetry can also be thought as the remnant of a Grand-Unified Theory (GUT) at higher energies. We have presented the results of a phenomenological analysis that explore the possibility to have doubly-charged bosonic resonances, with signature not allowed by the SM.

\end{document}